\documentclass[useAMS,usenatbib]{mnras}

\usepackage{graphicx}
\usepackage{txfonts}

\title[Peculiarities of chemical abundance distribution]
      {Peculiarities of the chemical abundance distribution in 
       galaxies NGC~3963 and NGC~7292}

\author[A.~S.~Gusev \& A.~V.~Dodin]
       {A.~S.~Gusev\thanks{E-mail:gusev@sai.msu.ru}  
        and A.~V.~Dodin \\
        Sternberg Astronomical Institute, Lomonosov Moscow State University, 
        Universitetsky pr. 13, 119234 Moscow, Russia \\
             }

\date{Accepted 2021 May 11. Received 2021 April 27; 
in original form 2021 April 24}

\begin{document}

\maketitle

\begin{abstract}
Spectroscopic observations of 32 H\,{\sc ii} regions in the spiral galaxy 
NGC~3963 and the barred irregular galaxy NGC~7292 were carried out with the 
2.5-m telescope of the Caucasus Mountain Observatory of the Sternberg 
Astronomical Institute using the Transient Double-beam Spectrograph with 
a dispersion of $\approx$1\AA\,pixel$^{-1}$ and a spectral resolution of 
$\approx$3\AA. These observations were used to estimate the oxygen and 
nitrogen abundances and the electron temperatures in H\,{\sc ii} regions 
through modern strong-line methods. In general, the galaxies have 
oxygen and nitrogen abundances typical of stellar systems with similar 
luminosities, sizes, and morphology. However, we have found some peculiarities 
in chemical abundance distributions in both galaxies. The distorted outer 
segment of the southern arm of NGC~3963 shows an excess oxygen and 
nitrogen abundances. Chemical elements abundances in NGC~7292 are 
constant and do not depend on the galactocentric distance. These 
peculiarities can be explained in terms of external gas accretion in the 
case of NGC~3963 and major merging for NGC~7292.
\end{abstract}


\section{Introduction}

The distribution of chemical elements in galaxies plays a key role 
in understanding their formation and evolution. Regions of ionized 
hydrogen with their bright emission spectra are good indicators 
of this distribution. The oxygen abundance is usually used as a 
tracer of current metallicity in spiral and irregular galaxies. 
Numerous spectroscopic observations of H\,{\sc ii} regions, beginning 
from \citet{searle1971}, reveal an exponential decrease in the 
oxygen-to-hydrogen ratio from the centre to the outer disc regions 
of galaxies.

Chemical elements gradients are the result of the galaxy evolution, 
where a complex interplay between star formation rate and efficiency, 
stellar migration, accretion and outflow of metal-poor 
and metal-rich gas, tidal interactions and mergers, as well as gas flows 
within the disc form the radial distribution of heavy chemical 
elements \citep{zurita2021}. The findings of investigations into variations 
of the gas composition within galaxies, in combination with results on the 
evolution of stellar populations, has led to the development of models of 
chemical evolution of galaxies \citep*{chiappini2003,marcon2010}.

Studies of possible relationships between 
the abundance properties and global characteristics of galaxies such 
as morphology and luminosity are equally important \citep{pagel1991,vilacostas1992,dutil1999,
sanchez2014,florido2015,perez2016,zinchenko2019,zurita2021b}.

Although the main indicator of metallicity is the oxygen abundance, 
the nitrogen-to-oxygen ratio plays an equally important role, being an 
indicator of star formation history \citep{mallery2007,molla2010}.

To determine parameters of the oxygen and nitrogen abundance 
distributions, the abundance estimates for H\,{\sc ii} regions, 
evenly distributed across the whole galaxy disc from the inner to the 
outer part, are necessary. This kind of measurements are available for 
a few hundred galaxies 
\citep*[see compilations in][]{pilyugin2014,zurita2021}.

Integral field unit spectroscopy instruments and large surveys, 
associated with them, provide the most complete data for chemical and 
kinematic analysis. However, a small field of view of such devices 
($74\times64$~arcsec$^2$ in CALIFA, for instance) confines the study 
of giant and nearby galaxies to their inner parts. Wide-field integral 
spectroscopy instruments provide observational data for smaller number 
of galaxies. We believe that classical long-slit and multi-slit 
spectroscopic data still play an important role, especially for studies 
of some individual galaxies with signs of peculiarity in morphology, 
which were not included in large spectral surveys.

\begin{figure*}
\resizebox{1.00\hsize}{!}{\includegraphics[angle=000]{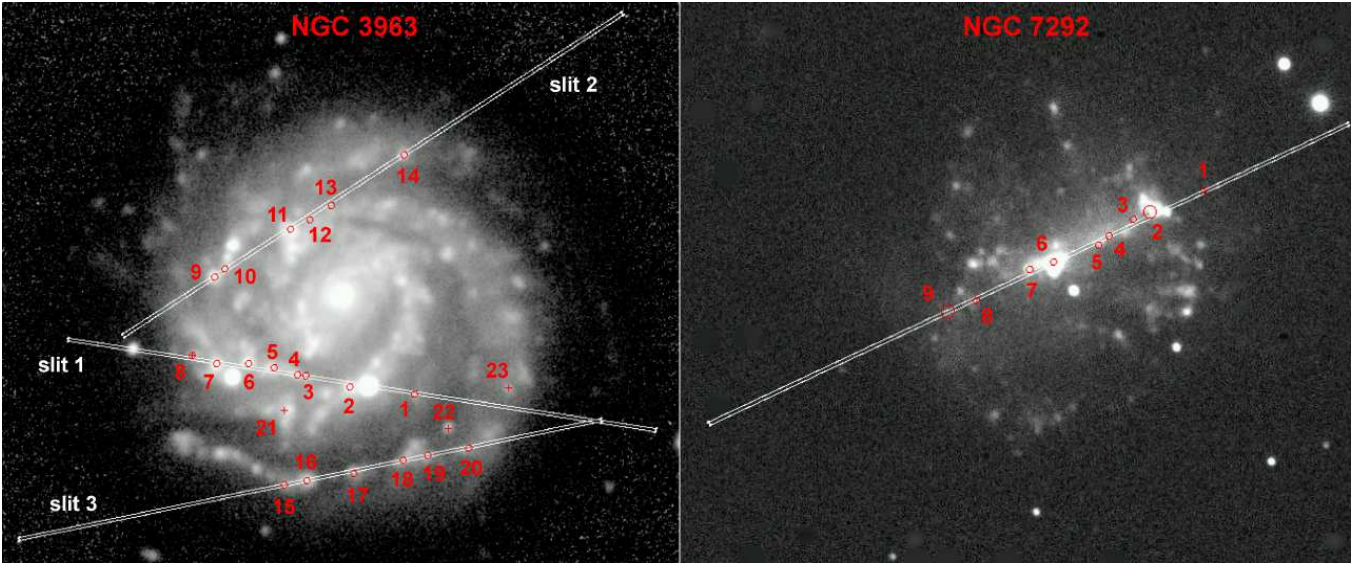}}
\caption{$U$ images of NGC~3963 (left) and NGC~7292 (right) with 
overlaid positions of the slits (white narrow bars). The sizes of 
the slits are $180\times1$ arcsec$^2$. The positions and numbers 
of the H\,{\sc ii} regions observed via TDS (open red circles) 
and obtained from the SDSS (red crosses) are indicated. The region no.~8 
in NGC~3963 (cross in circle) was observed both with the TDS and BOSS 
(SDSS). North is upwards and east is to the left.
}
\label{figure:map}
\end{figure*}

\begin{table*}
\caption[]{\label{table:sample}
The galaxy sample.
}
\begin{center}
\begin{tabular}{cccccccccccc} \hline \hline
Galaxy & Type & $B_t$ & $M_B^a$ & Inclination & PA       & $v^b$ & 
$R_{25}^c$ & $R_{25}^c$ & $d$   & $A(B)_{\rm Gal}$ & $A(B)_{\rm in}$ \\
       &      & (mag) & (mag)   & (degree)    & (degree) & (km\,s$^{-1}$) & 
(arcmin)   & (kpc)      & (Mpc) & (mag)        & (mag) \\
1 & 2 & 3 & 4 & 5 & 6 & 7 & 8 & 9 & 10 & 11 & 12 \\
\hline
NGC~3963 & SAB(rs)bc & $12.60\pm0.07$ & $-21.10\pm0.325$ & 
28.9 &  95.9 & 3184 & 1.285 & 18.39 & 49.2 & 0.083 & 0.09 \\
NGC~7292 & IBm       & $13.06\pm0.06$ & $-17.00\pm1.007$ & 
54.4 & 113.0 &  986 & 0.953 &  1.89 & 6.82 & 0.223 & 0.32 \\
\hline
\end{tabular}
\end{center}
\begin{flushleft}
$^a$ Absolute magnitude of a galaxy corrected for Galactic extinction and 
inclination effects. \\
$^b$ Heliocentric radial velocity. \\
$^c$ Radius of a galaxy at the isophotal level 25 mag\,arcsec$^{-2}$ in 
the $B$ band corrected for Galactic extinction and inclination effects. \\
\end{flushleft}
\end{table*}

The newly installed Transient Double-beam Spectrograph (TDS; see 
Section~\ref{sect:observ} for details) is already used actively in 
several large international projects such as the SRG/eROSITA All-Sky 
Survey \citep{dodin2020} and the Zwicky Transient Facility survey 
\citep{malanchev2021}, as well as in various stellar, galactic, 
and extragalactic studies 
\citep[see][and references therein]{potanin2020}. This paper presents 
the first results of a spectroscopic study of extragalactic 
H\,{\sc ii} regions obtained at the TDS.

We selected for spectroscopic study the Milky Way-type galaxy NGC~3963 
and NGC~7292 which is similar to Large Magellanic Cloud by size and 
morphology (Fig.~\ref{figure:map}, Table~\ref{table:sample}). Both 
galaxies host numerous H\,{\sc ii} regions. NGC~3963 is a rather 
symmetric grand-design galaxy with distorted outer part of southern 
spiral arm. It has a companion of comparable mass, NGC~3958, which is 
located at the distance of 7.7~arcmin (110~kpc) to the south from 
NGC~3963. Tidal distortions have been found in both galaxies of 
the pair \citep{moorsel1983}.

NGC~7292 is an irregular galaxy with a strong bright bar. The galaxy 
looks like a flower with three petals with the bar between two of them 
(Fig.~\ref{figure:map}). Let us remark that Magellanic-type galaxies is a rare 
galactic type. Data on the chemical elements abundance in H\,{\sc ii} 
regions in this type of stellar systems are known for a limited number 
of galaxies \citep[see e.g.][and references therein]{pilyugin2014}.

Both galaxies are not isolated and are influenced by the environment: 
gas accretion in the case of NGC~3963 and a major merging in the case of 
NGC~7292. The goal of our study is to check how the environment affects 
the chemical elements abundance and their spatial distribution in these 
galaxies.

Spectroscopic observations of H\,{\sc ii} regions in NGC~3963 and 
NGC~7292 have not been carried out previously, with the exception of 4 
regions, which were observed in the Sloan Digital Sky 
Survey\footnote{\url{https://www.sdss.org}} (SDSS). These regions 
are included in our study (see Fig.~\ref{figure:map}).

The fundamental parameters of the galaxies are presented in 
Table~\ref{table:sample}, where the morphological type, Galactic 
absorption, $A(B)_{\rm Gal}$, and the distance to NGC~3963 are taken from 
the NED\footnote{\url{http://ned.ipac.caltech.edu}} database, and the 
remaining parameters are taken from the 
LEDA\footnote{\url{http://leda.univ-lyon1.fr}} database. The distance 
to NGC~7292 is still an open question (see the distance estimates in NED). 
We use the value of the distance to NGC~7292 obtained in \citet{tully2009}.

\begin{figure*}
\resizebox{1.00\hsize}{!}{\includegraphics[angle=000]{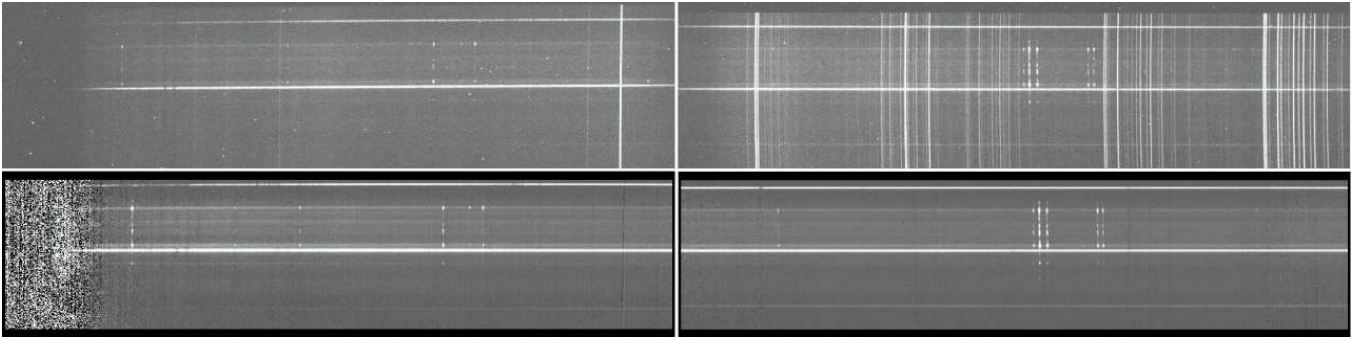}}
\caption{Frames with spectra of NGC~3963 obtained in the position of 
slit~1 in blue (left panels) and red (right panels) channels of 
the TDS. The raw spectra (top) and the spectra after initial reduction 
(bottom) are shown. The sizes of the images are $2048\times512$ (top) 
and $2030\times460$ pixels$^2$ (bottom).
}
\label{figure:spec_ima}
\end{figure*}

\section{Observations and data reduction}

\subsection{Observations}
\label{sect:observ}

The observations were carried out at the 2.5~m telescope of 
the Caucasus Mountain Observatory of the Sternberg Astronomical 
Institute (Lomonosov Moscow State University; CMO SAI MSU) with 
the TDS. This is a new device which was installed at the end of 
2019 \citep{potanin2020}.

\begin{table}
\caption[]{\label{table:observ}
Journal of observations.
}
\begin{center}
\begin{tabular}{ccccc} \hline \hline
Slit     & PA$_{\rm slit}$ & Date & Exposures & Air mass \\
position & (degree)        &      & (s)       &          \\
\hline
\multicolumn{5}{c}{NGC~3963} \\
1 &  81.2 & 2020.12.11/12 & $900\times4$     & 1.1 \\
2 & 123.0 & 2020.12.14/15 & $900\times3$     & 1.6 \\
3 & 101.7 & 2020.12.17/18 & $900\times6$     & 1.5 \\
\multicolumn{5}{c}{NGC~7292} \\
  & 118.3 & 2020.12.10/11 & $100+900\times3$ & 1.4 \\
\hline
\end{tabular}\\
\end{center}
\end{table}

The spectrograph operates simultaneously in blue (the range of 
3600--5770\AA\, with a dispersion of 1.21\AA\,pixel$^{-1}$ and a 
spectral resolution of 3.6\AA) and red (the range of 5673--7460\AA\, 
with a dispersion of 0.87\AA\,pixel$^{-1}$ and a spectral resolution 
of 2.6\AA) channels \citep{potanin2020}. Two CCD cameras use 
E2V~42-10 detectors with a size of $2048\times512$ pixels$^2$. 
The pixel size is 0.363~arcsec, the size of slits used is 
$180\times1$ arcsec$^2$. For a detailed description of the 
spectrograph, see \citet{potanin2020}.

Observations were carried out in December 2020 (see the journal of 
observations in Table~\ref{table:observ}). An example of raw spectrum 
obtained in both channels is presented in 
Fig.~\ref{figure:spec_ima} (top panel).

We chose such slit positions to cover the maximum number of H\,{\sc ii} 
regions in a wide range of galactocentric distances. As a result, 
we obtained spectra for 20 H\,{\sc ii} regions in NGC~3963 and 9 
H\,{\sc ii} regions in NGC~7292 (Fig.~\ref{figure:map}).

Spectral data for objects in NGC~3963 were obtained using three slit 
positions (H\,{\sc ii} regions nos.~1-8 were covered by the first, 
nos.~9-14 -- by the second, and nos.~15-20 -- by the third slit position; 
see Fig.~\ref{figure:map}).

Four H\,{\sc ii} regions in NGC~3963 have been previously observed in the 
SDSS with the BOSS spectrograph (nos. 8, 21-23). The region no.~8 in 
NGC~3963 was observed both on the TDS and BOSS (Fig.~\ref{figure:map}). 
1D spectra of these objects\footnote{Files: 
spec-8234-57450-0784.fits (no~8), spec-8237-58162-0466.fits (no~21), 
spec-8234-57450-0786.fits (no~22), and spec-8237-58162-0462.fits 
(no.~23)} were downloaded from the SDSS DR~16 
database.\footnote{\url{http://skyserver.sdss.org/dr16}}

\begin{figure}
\vspace{0.6cm}
\resizebox{1.00\hsize}{!}{\includegraphics[angle=000]{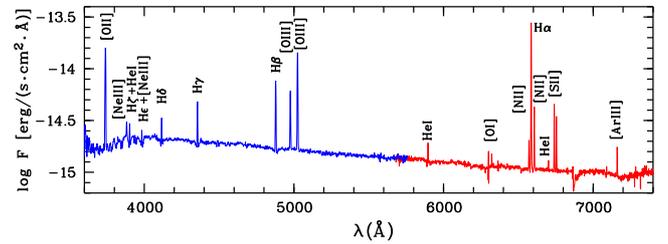}}
\caption{The spectrum of the H{\sc ii} region no.~6 (nucleus) of 
NGC~7292. The spectra obtained in the blue and red channels of the 
TDS are overlaid on one another.
}
\label{figure:spectrum}
\end{figure}

The seeing during our observations was from 1.2 up to 1.8~arcsec, typical 
angular diameters of studied H\,{\sc ii} regions are from 2 up to 
7~arcsec (Fig.~\ref{figure:map}). Remark that the last observational 
night was cloudy, as a result the spectra obtained in the slit position 
3 have a lower signal-to-noise ratio.

The observing procedure involved obtaining flat and wavelength 
calibration images at the beginning and end of each set. 
Spectrophotometric standards were observed immediately after the studied 
objects at the same air mass.

\begin{table*}
\caption[]{\label{table:flux1}
Parameters of H\,{\sc ii} regions, equivalent widths of the 
H$\alpha$ and H$\beta$ lines, and undereddened H$\beta$ line 
fluxes.
}
\begin{center}
\begin{tabular}{cccccccccc} \hline \hline
H\,{\sc ii} & N--S$^a$ & E--W$^a$ & $r^b$ & $r/R_{25}^b$ & 
$c$(H$\beta$) & $n_e$ & EW(H$\alpha$) & EW(H$\beta$) & $F$(H$\beta$) \\
region      & (arcsec) & (arcsec) & (kpc) & & & (cm$^{-3}$) & 
(\AA) & (\AA) & ($10^{-16}$ erg \,s$^{-1}$\,cm$^{-2}$) \\
\hline
\multicolumn{10}{c}{NGC~3963} \\
 1 & --30.0 &  +22.4 &  9.89 & 0.538 & $1.04\pm0.15$ & $<260$ & 
 $94.2\pm17.1$ & $18.7\pm3.9$ & $2.3\pm0.2$ \\
 2 & --27.8 &   +2.9 &  7.60 & 0.413 & $0.99\pm0.02$ & $<130$ & 
$185.1\pm12.7$ & $31.2\pm2.4$ & $29.6\pm0.4$ \\
 3 & --24.6 & --10.6 &  7.08 & 0.385 & $1.17\pm0.08$ & $<120$ & 
$105.2\pm10.4$ & $21.5\pm2.7$ & $3.8\pm0.2$ \\
 4 & --24.2 & --13.1 &  7.20 & 0.392 & $0.87\pm0.03$ & $<130$ & 
$155.6\pm13.9$ & $32.2\pm4.3$ & $9.1\pm0.2$ \\
 5 & --22.0 & --20.1 &  7.57 & 0.411 & $0.95\pm0.08$ & $<130$ & 
 $70.2\pm7.6$ & $16.0\pm1.5$ & $5.0\pm0.3$ \\
 6 & --20.8 & --28.0 &  8.63 & 0.469 & $1.30\pm0.27$ & $530\pm220$ & 
 $62.0\pm8.5$ & $14.2\pm1.9$ & $1.7\pm0.2$ \\
 7 & --20.8 & --37.5 & 10.46 & 0.569 & $0.92\pm0.02$ & $<60$ & 
$225.5\pm18.1$ & $33.1\pm2.9$ & $25.2\pm0.3$ \\
 8$^c$ & --18.2 & --44.8 & 11.67 & 0.635 & $0.83\pm0.30$ & $<110$ & 
$114.1\pm69.2$ & $20.9\pm12.2$ & $0.9\pm0.2$ \\
 8$^d$ & --18.2 & --44.8 & 11.67 & 0.635 & $-$ & $<110$ & 
 $60.9\pm22.7$ & $17.2\pm2.5$ & $1.8\pm0.1$ \\
 9 &   +5.6 & --38.1 &  9.28 & 0.504 & $1.12\pm0.05$ & $<35$ & 
$328.6\pm112.6$ & $42.3\pm8.7$ & $9.6\pm0.2$ \\
10 &   +8.1 & --35.0 &  8.70 & 0.473 & $0.88\pm0.08$ & $360\pm190$ & 
$217.5\pm99.8$ & $49.0\pm22.4$ & $4.6\pm0.2$ \\
11 &  +20.1 & --15.1 &  6.64 & 0.361 & $2.05\pm0.55$ & $<200$ & 
 $46.8\pm5.4$ &  $9.8\pm1.1$ & $1.4\pm0.3$ \\
12 &  +23.1 &  --9.2 &  6.71 & 0.365 & $1.59\pm0.52$ & $<580$ & 
 $61.9\pm15.0$ & $17.1\pm4.1$ & $1.2\pm0.3$ \\
13 &  +27.2 &  --2.8 &  7.46 & 0.406 & $1.90\pm0.92$ & $<360$ & 
 $63.5\pm17.7$ & $15.7\pm5.5$ & $0.8\pm0.3$ \\
14 &  +43.0 &  +19.4 & 12.46 & 0.678 & $1.05\pm0.12$ & $<120$ & 
$214.1\pm104.6$ & $35.5\pm16.0$ & $4.6\pm0.3$ \\
15 & --57.7 & --17.0 & 16.10 & 0.876 & $1.74\pm1.37$ & $<660$ & 
 $60.6\pm23.1$ & $13.3\pm5.1$ & $3.0\pm1.6$ \\
16 & --56.3 & --10.2 & 15.44 & 0.840 & $2.01\pm1.84$ & $<870$ & 
 $46.5\pm9.9$ & $12.2\pm2.9$ & $2.9\pm1.9$ \\
17 & --54.0 &   +4.1 & 14.76 & 0.803 & $1.04\pm0.10$ & $<180$ & 
$187.9\pm64.2$ & $32.3\pm9.7$ & $32.8\pm1.8$ \\
18 & --50.2 &  +18.9 & 14.51 & 0.789 & $0.92\pm0.09$ & $<130$ & 
$125.7\pm26.6$ & $25.3\pm5.3$ & $28.6\pm1.6$ \\
19 & --48.7 &  +26.4 & 14.82 & 0.806 & $0.85\pm0.14$ & $<360$ & 
 $90.8\pm23.3$ & $21.7\pm5.5$ & $13.8\pm1.4$ \\
20 & --46.5 &  +38.9 & 15.89 & 0.864 & $1.20\pm0.19$ & $<250$ & 
 $67.6\pm8.9$ & $18.6\pm3.5$ & $16.3\pm1.6$ \\
21 & --35.0 & --16.9 & 10.24 & 0.557 & $-$ & $<110$ & 
 $34.4\pm3.5$ & $14.8\pm1.7$ & $1.5\pm0.2$ \\
22 & --40.4 &  +32.5 & 13.64 & 0.742 & $-$ & $<85$ & 
$194.4\pm28.9$ & $55.0\pm9.3$ & $11.0\pm0.1$ \\
23 &  -28.2 &  +51.0 & 14.56 & 0.792 & $-$ & $<100$ & 
 $39.4\pm5.7$ & $13.5\pm2.2$ & $1.5\pm0.1$ \\
\multicolumn{10}{c}{NGC~7292} \\
 1 &  +18.5 &  +38.6 &  1.42 & 0.750 & $0.84\pm0.20$ & $<120$ & 
 $59.3\pm12.1$ & $19.9\pm5.7$ & $2.9\pm0.4$ \\
 2 &  +12.9 &  +24.7 &  0.93 & 0.490 & $0.68\pm0.01$ & $<40$ & 
$152.9\pm6.0$  & $28.0\pm1.2$ & $108.7\pm0.9$ \\
 3 &  +11.0 &  +20.3 &  0.77 & 0.407 & $0.78\pm0.02$ & $<65$ & 
 $98.9\pm5.8$  & $23.6\pm1.4$ & $31.9\pm0.6$ \\
 4 &   +6.7 &  +14.3 &  0.52 & 0.277 & $1.12\pm0.11$ & $<40$ & 
 $48.9\pm2.9$  & $15.5\pm1.0$ & $9.2\pm0.6$ \\
 5 &   +4.5 &  +11.5 &  0.41 & 0.217 & $0.63\pm0.05$ & $<45$ & 
 $41.3\pm2.0$  & $14.7\pm0.9$ & $10.9\pm0.6$ \\
 6 &    0.0 &    0.0 &  0.00 & 0.000 & $0.57\pm0.01$ & $<60$ & 
$117.3\pm3.5$  & $26.0\pm0.9$ & $226.4\pm2.0$ \\
 7 &  --1.8 &  --6.1 &  0.21 & 0.113 & $1.03\pm0.04$ & $<20$ & 
$131.8\pm7.5$  & $25.1\pm2.3$ & $42.9\pm0.9$ \\
 8 &  --9.7 & --19.5 &  0.72 & 0.382 & $0.98\pm0.11$ & $<30$ & 
 $79.1\pm10.3$ & $17.8\pm2.6$ & $9.2\pm0.7$ \\
 9 & --12.8 & --26.9 &  0.99 & 0.522 & $0.62\pm0.06$ & $<20$ & 
 $80.9\pm13.6$ & $22.0\pm3.4$ & $12.6\pm0.7$ \\
\hline
\end{tabular}
\end{center}
\begin{flushleft}
$^a$ Offsets from the galactic centre, positive to the north and west. \\
$^b$ Deprojected galactocentric distance. \\
$^c$ From TDS spectrum. \\ $^d$ From SDSS spectrum.
\end{flushleft}
\end{table*}

\subsection{Data reduction}

Initial data reduction followed routine procedures, including dark, 
cosmic-ray, and flat-field corrections, wavelength calibration with a 
standard Ne-Al-Si lamp, rebinning the data onto a uniform wavelength grid, 
wavelength correction using night sky lines, photometric calibration, 
summation of spectra, subtracting the background, and transformation to 
one-dimensional spectrum \citep[see][for details]{potanin2020}. To reduce 
raw spectra, we used a {\sc python}-based data processing package, developed 
in SAI MSU, and the European Southern Observatory Munich Image Data Analysis 
System ({\sc eso-midas}). An example of final 2D spectrum is 
presented in Fig.~\ref{figure:spec_ima} (bottom panel). An example of
1D-spectrum after reduction is shown in Fig.~\ref{figure:spectrum}.

The emission line fluxes were measured using the continuum-subtracted 
spectrum. Flux calibration was performed, using standard stars 
BD+75$\degr$325, HZ21, and HR4554 from the ESO 
list.\footnote{\url{https://www.eso.org/sci/observing/tools/standards.html}}

The extraction aperture corresponded to the area, where the brightest 
emission lines from H\,{\sc ii} regions were ''visible'' above the noise. 
This size is approximately equal to the angular diameter of individual 
H\,{\sc ii} regions projected along the PA of the slit. Thus the apparent 
sizes of the studied H\,{\sc ii} regions are larger than the width of the 
slit.

Coordinates, deprojected galactocentric distances of H\,{\sc ii} regions, 
the logarithmic extinction coefficient $c$(H$\beta$), equivalent widths of 
the  H$\alpha$ and H$\beta$ lines, and undereddened H$\beta$ line fluxes are 
listed in Table~\ref{table:flux1}. We also give in this table the electron 
densities, $n_e$, of H\,{\sc ii} regions obtained from the 
[S\,{\sc ii}]$\lambda$6717/[S\,{\sc ii}]$\lambda$6731 ratio according 
to \citet*{proxauf2014}.

When calculating the error of the line flux measurement, we took into 
account the following factors: the Poisson statistics of the line photon 
flux, the error connected with the computation of the underlying continuum, 
and the uncertainty of the spectral sensitivity curve. All these components are 
summed in quadrature. The total errors have been propagated to calculate the 
errors of all derived parameters. The use of the 1~arcsec slit introduces an 
unknown factor up to~2 due to slit losses, which, however, is constant within 
3~percent over the spectral range.

\begin{table*}
\caption[]{\label{table:flux2} The reddening-corrected fluxes of main 
emission lines of H\,{\sc ii} regions in units of $I({\rm H}\beta)$.}
\begin{center}
\begin{tabular}{cccccccc} \hline \hline
H\,{\sc ii} & [O\,{\sc ii}] & [O\,{\sc iii}] & [O\,{\sc iii}] & 
[N\,{\sc ii}] & [N\,{\sc ii}] & [S\,{\sc ii}] & [S\,{\sc ii}] \\
region & 3727+3729 & 4959 & 5007 & 6548 & 6584 & 6717 & 6731 \\
\hline
\multicolumn{8}{c}{NGC~3963} \\
 1 & $5.88\pm1.10$ & $0.17\pm0.09$ & $0.50\pm0.10$ & $0.24\pm0.05$ & 
$0.78\pm0.13$ & $0.40\pm0.07$ & $0.30\pm0.06$ \\
 2 & $2.21\pm0.10$ & $0.06\pm0.01$ & $0.23\pm0.01$ & $0.30\pm0.01$ & 
$0.95\pm0.02$ & $0.36\pm0.01$ & $0.27\pm0.01$ \\
 3 & $4.02\pm0.39$ & $0.11\pm0.04$ & $0.33\pm0.05$ & $0.34\pm0.03$ & 
$1.06\pm0.09$ & $0.42\pm0.04$ & $0.30\pm0.03$ \\
 4 & $2.49\pm0.13$ & $0.10\pm0.02$ & $0.23\pm0.02$ & $0.33\pm0.01$ & 
$1.03\pm0.04$ & $0.39\pm0.02$ & $0.29\pm0.01$ \\
 5 & $3.36\pm0.42$ & $-$           & $0.17\pm0.05$ & $0.32\pm0.03$ & 
$0.99\pm0.09$ & $0.49\pm0.05$ & $0.35\pm0.04$ \\
 6 & $8.44\pm2.43$ & $-$           & $0.52\pm0.13$ & $0.28\pm0.09$ & 
$0.88\pm0.25$ & $0.49\pm0.15$ & $0.51\pm0.15$ \\
 7 & $3.53\pm0.10$ & $0.28\pm0.02$ & $0.78\pm0.02$ & $0.23\pm0.01$ & 
$0.75\pm0.02$ & $0.42\pm0.01$ & $0.29\pm0.01$ \\
 8 & $7.23\pm0.55$ & $0.13\pm0.04$ & $0.78\pm0.08$ & $0.28\pm0.03$ & 
$0.82\pm0.06$ & $0.67\pm0.05$ & $0.46\pm0.04$ \\
 9 & $3.03\pm0.15$ & $0.09\pm0.02$ & $0.45\pm0.02$ & $0.32\pm0.02$ & 
$0.84\pm0.04$ & $0.52\pm0.03$ & $0.34\pm0.02$ \\
10 & $2.96\pm0.25$ & $0.12\pm0.04$ & $0.35\pm0.05$ & $0.32\pm0.04$ & 
$0.89\pm0.08$ & $0.36\pm0.04$ & $0.33\pm0.04$ \\
11 & $14.71\pm7.55$ & $-$           & $-$           & $-$           & 
$0.95\pm0.52$ & $0.49\pm0.28$ & $0.34\pm0.20$ \\
12 & $5.49\pm2.74$ & $-$           & $-$           & $0.34\pm0.18$ & 
$1.07\pm0.54$ & $0.49\pm0.26$ & $0.41\pm0.22$ \\
14 & $4.50\pm0.56$ & $0.24\pm0.06$ & $0.73\pm0.08$ & $0.30\pm0.04$ & 
$0.81\pm0.10$ & $0.51\pm0.07$ & $0.34\pm0.05$ \\
17 & $4.49\pm0.50$ & $0.34\pm0.06$ & $0.70\pm0.06$ & $0.25\pm0.03$ & 
$0.78\pm0.08$ & $0.56\pm0.07$ & $0.41\pm0.05$ \\
18 & $5.02\pm0.49$ & $0.37\pm0.06$ & $0.80\pm0.07$ & $0.30\pm0.03$ & 
$0.80\pm0.08$ & $0.61\pm0.06$ & $0.42\pm0.05$ \\
19 & $5.01\pm0.85$ & $0.39\pm0.10$ & $0.55\pm0.11$ & $0.35\pm0.07$ & 
$0.93\pm0.15$ & $0.73\pm0.13$ & $0.60\pm0.11$ \\
20 & $6.68\pm1.39$ & $0.36\pm0.12$ & $0.85\pm0.14$ & $0.32\pm0.07$ & 
$0.80\pm0.15$ & $0.52\pm0.11$ & $0.38\pm0.08$ \\
21 & $7.75\pm1.18$ & $-$           & $0.29\pm0.09$ & $0.30\pm0.05$ & 
$0.96\pm0.11$ & $0.68\pm0.09$ & $0.44\pm0.07$ \\
22 & $2.40\pm0.07$ & $0.26\pm0.01$ & $0.74\pm0.01$ & $0.20\pm0.01$ & 
$0.60\pm0.01$ & $0.32\pm0.01$ & $0.23\pm0.01$ \\
23 & $6.85\pm0.79$ & $0.31\pm0.08$ & $0.83\pm0.10$ & $0.31\pm0.08$ & 
$0.94\pm0.09$ & $0.84\pm0.08$ & $0.57\pm0.06$ \\
\multicolumn{8}{c}{NGC~7292} \\
 1 & $10.63\pm2.43$ & $1.37\pm0.26$ & $4.53\pm0.69$ & $-$           & 
$0.38\pm0.10$ & $0.72\pm0.17$ & $0.46\pm0.12$ \\
 2 & $5.09\pm0.07$ & $0.86\pm0.01$ & $2.55\pm0.02$ & $0.09\pm0.00$ & 
$0.29\pm0.00$ & $0.42\pm0.01$ & $0.29\pm0.00$ \\
 3 & $5.41\pm0.16$ & $0.74\pm0.02$ & $2.07\pm0.04$ & $0.11\pm0.01$ & 
$0.42\pm0.01$ & $0.55\pm0.02$ & $0.39\pm0.01$ \\
 4 & $12.46\pm1.43$ & $0.85\pm0.07$ & $2.25\pm0.16$ & $0.18\pm0.03$ & 
$0.54\pm0.07$ & $0.82\pm0.10$ & $0.55\pm0.07$ \\
 5 & $8.65\pm0.63$ & $0.47\pm0.05$ & $1.47\pm0.09$ & $0.12\pm0.02$ & 
$0.52\pm0.04$ & $0.75\pm0.06$ & $0.50\pm0.04$ \\
 6 & $4.14\pm0.05$ & $0.70\pm0.01$ & $2.08\pm0.02$ & $0.10\pm0.00$ & 
$0.32\pm0.00$ & $0.32\pm0.00$ & $0.23\pm0.00$ \\
 7 & $5.80\pm0.23$ & $0.64\pm0.02$ & $2.03\pm0.05$ & $0.12\pm0.01$ & 
$0.36\pm0.01$ & $0.50\pm0.02$ & $0.32\pm0.01$ \\
 8 & $6.92\pm0.86$ & $0.46\pm0.06$ & $1.50\pm0.12$ & $0.17\pm0.03$ & 
$0.44\pm0.06$ & $0.67\pm0.09$ & $0.42\pm0.06$ \\
 9 & $5.74\pm0.48$ & $0.71\pm0.06$ & $1.75\pm0.11$ & $0.14\pm0.03$ & 
$0.42\pm0.04$ & $0.52\pm0.04$ & $0.31\pm0.03$ \\
\hline
\end{tabular}
\end{center}
\end{table*}

The measured emission fluxes $F$ were corrected for the interstellar 
reddening and Balmer absorption in the underlying stellar continuum. 
The theoretical H$\alpha$/H$\beta$ ratio from \citet{osterbrock1989}, assuming 
case B recombination and an electron temperature of 10$^4$~K and the 
analytical approximation to the Whitford interstellar reddening law by 
\citet*{izotov1994}, were used. We adopted the absorption equivalent width of 
hydrogen lines EW$_a$($\lambda$)=2\AA\, according to \citet*{mccall1985}, 
and EW$_a$($\lambda$)=0 for lines other than hydrogen.
 
Reddening-corrected line intensities $I$($\lambda$)/$I$(H$\beta$) are given 
in Table~\ref{table:flux2}.

\section{Results}

\subsection{Preliminary remarks}
\label{sect:pre_res}

\begin{figure}
\vspace{0.9cm}
\hspace{0.7cm}
\resizebox{0.76\hsize}{!}{\includegraphics[angle=000]{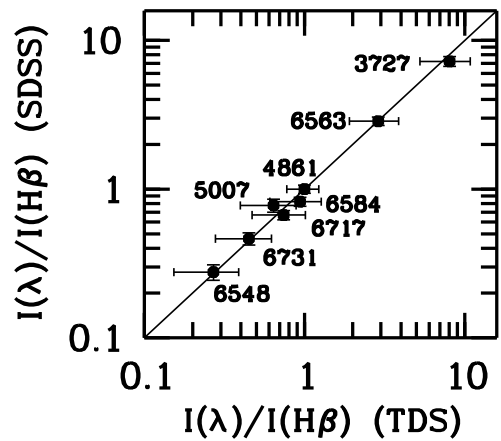}}
\caption{Comparison between reddening-corrected fluxes for the region 
no~8 in NGC~3963 obtained from the TDS and SDSS spectra. 
Corresponding wavelengths of spectral lines in \AA\, are given. 
The error crosses are shown.
}
\label{figure:compare}
\end{figure}

As we noted above, the H\,{\sc ii} region no.~8 in NGC~3963 was observed with 
both TDS and BOSS. We compared our measurements from the TDS spectrum 
with results from the BOSS spectrum in Fig.~\ref{figure:compare}. As seen 
from the figure, the fluxes, obtained from the TDS and BOSS spectra, 
coincide within the error limits for all main emission lines.

At the same time, the fluxes obtained in the BOSS spectrum, turned out to 
be $1.9\pm0.2$ times higher than in the TDS spectrum. This is obviously 
due to the different apertures used. The BOSS has a 3~arcsec round 
aperture, while on the TDS a 1~arcsec width slit is used.

The measurement errors for emission lines, obtained from the BOSS spectrum, 
turned out to be significantly smaller than those obtained from the TDS 
spectrum. This is due to the fact that object no.~8 is one of two 
faintest objects in our sample, so it has relatively large errors in measured 
fluxes and $c$(H$\beta$) (see Table~\ref{table:flux2}). An additional 
factor is that the SDSS spectra have already been corrected for reddening, 
so we do not take into account their $c$(H$\beta$) errors. Further we 
will use the results of measurements of the region no.~8 in NGC~3963, 
obtained from the BOSS spectrum.

As is known, emission spectra can be created by different excitation 
mechanisms. We examined the studied H\,{\sc ii} regions on the emission-line 
diagnostic diagram [N\,{\sc ii}]$\lambda$6584/H$\alpha$ versus 
[O\,{\sc iii}]$\lambda$5007/H$\beta$ \citep*{baldwin1981}.

\begin{figure}
\vspace{0.8cm}
\hspace{0.7cm}
\resizebox{0.80\hsize}{!}{\includegraphics[angle=000]{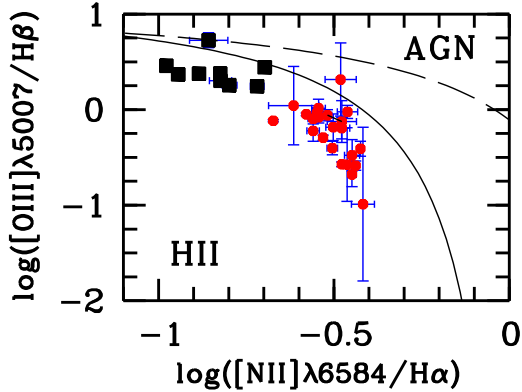}}
\caption{Emission-line diagnostic diagram for H{\sc ii} regions in 
NGC~3963 (small red circles) and NGC~7292 (large black squares). 
The curves separate objects with photoionized spectra from the 
objects with nonthermal emission spectra (AGN) according to 
\citet{kauffmann2003} (solid curve) and \citet{kewley2001} 
(dashed curve). The error crosses are shown.
}
\label{figure:nhoh}
\end{figure}

Figure~\ref{figure:nhoh} shows that most of H\,{\sc ii} regions from our 
sample are definitely thermally photoionized objects, that is, classical 
H\,{\sc ii} regions. Three regions of our sample (no.~16 in NGC~3963 and 
nos.~1 and 4 in NGC~7292) are located near the border separating objects 
with thermal and non-thermal emissions. These objects will be discussed 
in Section~\ref{sect:discus3}. Nevertheless, all these H\,{\sc ii} regions 
are included in further consideration. 

It should be noted that the samples of H\,{\sc ii} regions in galaxies 
NGC~3963 and NGC~7292 are located on the diagram in Fig.~\ref{figure:nhoh} 
separately from each other. Objects in NGC~3963 have a systematically lower 
[O\,{\sc iii}]$\lambda$5007/H$\beta$ ratio and a higher 
[N\,{\sc ii}]$\lambda$6584/H$\alpha$ ratio than those in NGC~7292.

We eliminated from further consideration H\,{\sc ii} regions with large 
errors in the measured line fluxes. Basically, these errors are the result 
of large uncertainties in estimates of the extinction coefficient $c$(H$\beta$). 
The flux measurement errors are comparable to the fluxes themselves for all 
the main emission lines in regions nos.~13, 15 and 16 in NGC~3963, in which 
$\Delta c$(H$\beta)>0.9$ (Table~\ref{table:flux1}). These H\,{\sc ii} 
regions are not included in Table~\ref{table:flux2}.

Extinction coefficients in H\,{\sc ii} regions nos.~11 and 12 in NGC~3963 
were estimated with errors of $>0.5$ (Table~\ref{table:flux1}). As a result, 
the line fluxes were measured with an accuracy of $\approx50\%$ 
(Table~\ref{table:flux2}). Moreover, we were unable to measure 
[O\,{\sc iii}]$\lambda$4959 and [O\,{\sc iii}]$\lambda$5007 lines for these 
objects. These two regions were also excluded from further analysis.

\subsection{Oxygen and nitrogen abundances}

We are unable to use the direct $T_e$-method to estimate the oxygen and nitrogen 
abundances, since auroral lines in spectra of H\,{\sc ii} regions from our 
sample, such as [O\,{\sc iii}]$\lambda$4363 or [N\,{\sc ii}]$\lambda$5755, 
are too faint to be detected. Thus, we need to use empirical calibrations. 
Such methods are well developed \citep*{kobul2004,pettini2004,pilyugin2005,
bresolin2007,pilyugin2010,pilyugin2011,marino2013,pilyugin2016}.

\begin{figure}
\vspace{0.9cm}
\hspace{0.7cm}
\resizebox{0.80\hsize}{!}{\includegraphics[angle=000]{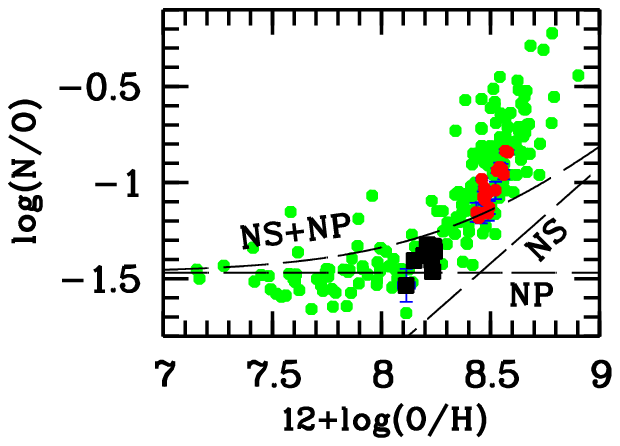}}
\caption{The N/O--O/H diagram for the investigated H\,{\sc ii} regions 
in NGC~3963 and NGC~7292. The green circles show $T_e$-based 
abundances in the sample of best-studied H\,{\sc ii} regions in nearby 
galaxies from the compilation of data of \citet{pilyugin2010} with 
additional data from \citet{gusev2012} and \citet*{gusev2013}. 
The dashed lines show possible boundaries for data points on the 
N/O--O/H plane under the assumption of a closed-box model for primary 
(NP), secondary (NS) and both primary and secondary (NS+NP) nitrogen 
according to \citet{vilacostas1993}. Chemical abundances of H\,{\sc ii} 
regions of our sample based on R-calibration are shown. Other symbols 
are the same as in Fig.~\ref{figure:nhoh}.
}
\label{figure:noh}
\end{figure}

The most modern two-dimensional R and S-calibrations \citep{pilyugin2016} 
are the simplest and the most popular now. The R-calibration, based on the 
[O\,{\sc ii}] and [N\,{\sc ii}] strong emission lines, is more resistant to 
the presence of diffuse ionized gas 
\citep[see e.g.][]{sanders2017,kumari2019,poetrodjojo2019}, but it is 
sensitive to extinction errors. The S-method, based on the [N\,{\sc ii}] 
and [S\,{\sc ii}] lines, does not depend strongly on extinction, but it 
can not be used for the nitrogen abundance estimates \citep{pilyugin2016}. 
Among the other methods we distinguish the NS-calibration \citep{pilyugin2011}, 
which is based on the [O\,{\sc iii}], [N\,{\sc ii}], and [S\,{\sc ii}] 
emission lines intensities 
\citep*[see discussion in][]{gusev2012,gusev2014}.

All the empirical calibrations are constructed under some assumptions. One 
of them is that the H\,{\sc ii} regions are in the low-density regime 
\citep{pilyugin2011,pilyugin2016}, which is typical for the majority of 
extragalactic H\,{\sc ii} regions 
\citep*{zaritsky1994,bresolin2005,gutierrez2010}.

The dependence $n_e$ versus 
[S\,{\sc ii}]$\lambda$6717/[S\,{\sc ii}]$\lambda$6731 degenerates at low 
$n_e$ \citep{proxauf2014}, thus we give upper limits of $n_e$ in 
Table~\ref{table:flux1} for most H\,{\sc ii} regions. Large upper 
limits of $n_e$ for H\,{\sc ii} regions nos.~12, 15, and 16 in NGC~3963 
are rather a result of large line flux errors. However, one H\,{\sc ii} 
region of our sample, no.~6 in NGC~3963, is an apparently high-density 
object (see Table~\ref{table:flux1}). The empirical calibrations are not 
applicable to such regions. As a result, this region was eliminated from 
further consideration.

We calculated the oxygen and nitrogen abundances in H\,{\sc ii} regions 
using three different calibrations discussed above. Data of emission lines 
fluxes from Table~\ref{table:flux2} were used. For several regions, where 
the [O\,{\sc iii}]$\lambda$4959 or [N\,{\sc ii}]$\lambda$6548 lines fluxes 
have not been measured, we adopt 
\begin{center}
$I$([O\,{\sc iii}]$\lambda$4959+[O\,{\sc iii}]$\lambda5007)=
1.33I$([O\,{\sc iii}]$\lambda$5007) 
\end{center}
and 
\begin{center}
$I$([N\,{\sc ii}]$\lambda$6548+[N\,{\sc ii}]$\lambda6584)=
1.33I$([N\,{\sc ii}]$\lambda$6584) 
\end{center} 
according to results of \citet{storey2000}.

\begin{table*}
\caption[]{\label{table:abun}
Oxygen and nitrogen abundances and electron temperatures in 
H\,{\sc ii} regions derived using the R, NS, and S calibrations.
}
\begin{center}
\begin{tabular}{cccccccc} \hline \hline
H\,{\sc ii} region & $r/R_{25}$ & \multicolumn{3}{c}{12+log(O/H)} & 
\multicolumn{2}{c}{12+log(N/H)} & $T_e$ \\
Calibration & & R & NS & S & R & NS & NS \\
\hline
\multicolumn{8}{c}{NGC~3963} \\
 1 & 0.538 & $8.44\pm0.02$ & $8.53\pm0.01$ & $8.49\pm0.02$ & 
$7.28\pm0.05$ & $7.62\pm0.03$ & $0.77\pm0.02$ \\
 2 & 0.413 & $8.57\pm0.00$ & $8.61\pm0.00$ & $8.56\pm0.00$ & 
$7.73\pm0.01$ & $7.89\pm0.00$ & $0.69\pm0.01$ \\
 3 & 0.385 & $8.56\pm0.01$ & $8.56\pm0.01$ & $8.58\pm0.01$ & 
$7.60\pm0.02$ & $7.83\pm0.02$ & $0.73\pm0.02$ \\
 4 & 0.392 & $8.58\pm0.00$ & $8.59\pm0.00$ & $8.58\pm0.00$ & 
$7.74\pm0.01$ & $7.88\pm0.01$ & $0.70\pm0.01$ \\
 5 & 0.411 & $8.55\pm0.01$ & $8.60\pm0.01$ & $8.54\pm0.01$ & 
$7.62\pm0.03$ & $7.84\pm0.02$ & $0.69\pm0.03$ \\
 7 & 0.569 & $8.47\pm0.00$ & $8.49\pm0.00$ & $8.48\pm0.00$ & 
$7.44\pm0.01$ & $7.52\pm0.00$ & $0.82\pm0.00$ \\
 8 & 0.635 & $8.45\pm0.01$ & $8.44\pm0.00$ & $8.46\pm0.01$ & 
$7.26\pm0.02$ & $7.42\pm0.01$ & $0.84\pm0.01$ \\
 9 & 0.504 & $8.53\pm0.00$ & $8.52\pm0.00$ & $8.51\pm0.01$ & 
$7.59\pm0.01$ & $7.64\pm0.01$ & $0.77\pm0.01$ \\
10 & 0.473 & $8.54\pm0.01$ & $8.56\pm0.01$ & $8.54\pm0.01$ & 
$7.62\pm0.02$ & $7.76\pm0.02$ & $0.74\pm0.01$ \\
14 & 0.678 & $8.49\pm0.01$ & $8.47\pm0.01$ & $8.49\pm0.02$ & 
$7.43\pm0.03$ & $7.53\pm0.02$ & $0.82\pm0.01$ \\
17 & 0.803 & $8.47\pm0.01$ & $8.45\pm0.01$ & $8.46\pm0.01$ & 
$7.39\pm0.03$ & $7.44\pm0.02$ & $0.84\pm0.01$ \\
18 & 0.789 & $8.47\pm0.01$ & $8.43\pm0.01$ & $8.47\pm0.01$ & 
$7.38\pm0.02$ & $7.42\pm0.02$ & $0.86\pm0.01$ \\
19 & 0.806 & $8.52\pm0.02$ & $8.42\pm0.01$ & $8.49\pm0.02$ & 
$7.48\pm0.04$ & $7.42\pm0.03$ & $0.86\pm0.01$ \\
20 & 0.864 & $8.46\pm0.02$ & $8.44\pm0.01$ & $8.49\pm0.02$ & 
$7.30\pm0.05$ & $7.47\pm0.03$ & $0.85\pm0.01$ \\
21 & 0.557 & $8.49\pm0.01$ & $8.52\pm0.01$ & $8.50\pm0.02$ & 
$7.33\pm0.04$ & $7.63\pm0.03$ & $0.76\pm0.03$ \\
22 & 0.742 & $8.46\pm0.00$ & $8.53\pm0.00$ & $8.45\pm0.00$ & 
$7.48\pm0.01$ & $7.55\pm0.00$ & $0.79\pm0.00$ \\
23 & 0.792 & $8.50\pm0.01$ & $8.39\pm0.01$ & $8.48\pm0.02$ & 
$7.37\pm0.03$ & $7.36\pm0.02$ & $0.89\pm0.01$ \\
\multicolumn{8}{c}{NGC~7292} \\
1 & 0.750 & $8.11\pm0.04$ & $8.30\pm0.02$ & $8.22\pm0.04$ & 
$6.58\pm0.08$ & $6.82\pm0.05$ & $1.07\pm0.01$ \\
2 & 0.490 & $8.15\pm0.00$ & $8.26\pm0.00$ & $8.21\pm0.00$ & 
$6.75\pm0.00$ & $6.84\pm0.00$ & $1.00\pm0.00$ \\
3 & 0.407 & $8.24\pm0.00$ & $8.31\pm0.00$ & $8.26\pm0.00$ & 
$6.91\pm0.01$ & $6.91\pm0.01$ & $0.97\pm0.00$ \\
4 & 0.277 & $8.23\pm0.02$ & $8.36\pm0.01$ & $8.31\pm0.02$ & 
$6.77\pm0.03$ & $6.96\pm0.02$ & $0.96\pm0.01$ \\
5 & 0.217 & $8.23\pm0.01$ & $8.32\pm0.01$ & $8.28\pm0.01$ & 
$6.83\pm0.02$ & $6.90\pm0.01$ & $0.93\pm0.00$ \\
6 & 0.000 & $8.21\pm0.00$ & $8.31\pm0.00$ & $8.27\pm0.00$ & 
$6.89\pm0.00$ & $6.97\pm0.00$ & $0.95\pm0.00$ \\
7 & 0.113 & $8.19\pm0.00$ & $8.30\pm0.00$ & $8.25\pm0.01$ & 
$6.81\pm0.01$ & $6.90\pm0.01$ & $0.96\pm0.00$ \\
8 & 0.382 & $8.24\pm0.01$ & $8.32\pm0.01$ & $8.28\pm0.02$ & 
$6.89\pm0.03$ & $6.92\pm0.02$ & $0.93\pm0.01$ \\
9 & 0.522 & $8.24\pm0.01$ & $8.33\pm0.01$ & $8.29\pm0.01$ & 
$6.91\pm0.03$ & $6.98\pm0.02$ & $0.94\pm0.01$ \\
\hline
\end{tabular}
\end{center}
\end{table*}

The resultant oxygen abundances from the R, NS, and S-calibrations 
abundances, nitrogen abundances from the R and NS-calibrations, and 
NS-calibration electron temperatures are given in Table~\ref{table:abun}.

Chemical elements abundance estimates obtained using various calibrations 
give small systematic shifts, $\sim0.1$~dex for oxygen and $\sim0.15$~dex 
for nitrogen. These are typical shifts between different empirical 
calibrations \citep[see e.g.][and references therein]{kewley2008,zurita2021}. 
Remark that the intrinsic accuracy of strong lines methods is equal to 
0.05-0.1~dex \citep{pilyugin2010,pilyugin2011,pilyugin2016}. These errors 
are not included in errors in Table~\ref{table:abun}.

Obtained oxygen and nitrogen abundances and electron temperatures are 
typical of galaxies with similar luminosities \citep[see e.g.][]{zurita2021b}. 
The O/H--N/O diagram (Fig.~\ref{figure:noh}) illustrates this well. All our 
points lie within the spread of sample of best-studied H\,{\sc ii} regions 
in nearby galaxies, where the chemical elements abundances were derived using the
direct $T_e$-method. The diagram shows that NGC~3963 is a system that is rich in 
oxygen and nitrogen, in which secondary nitrogen dominates. The nitrogen abundance 
in H\,{\sc ii} regions of NGC~3963 increases at a faster rate than the oxygen 
abundance. H\,{\sc ii} regions in NGC~7292 have lower oxygen abundance and 
a domination of primary nitrogen (Fig.~\ref{figure:noh}).

\subsection{Radial abundance gradients}

\begin{figure*}
\vspace{0.8cm}
\resizebox{0.92\hsize}{!}{\includegraphics[angle=000]{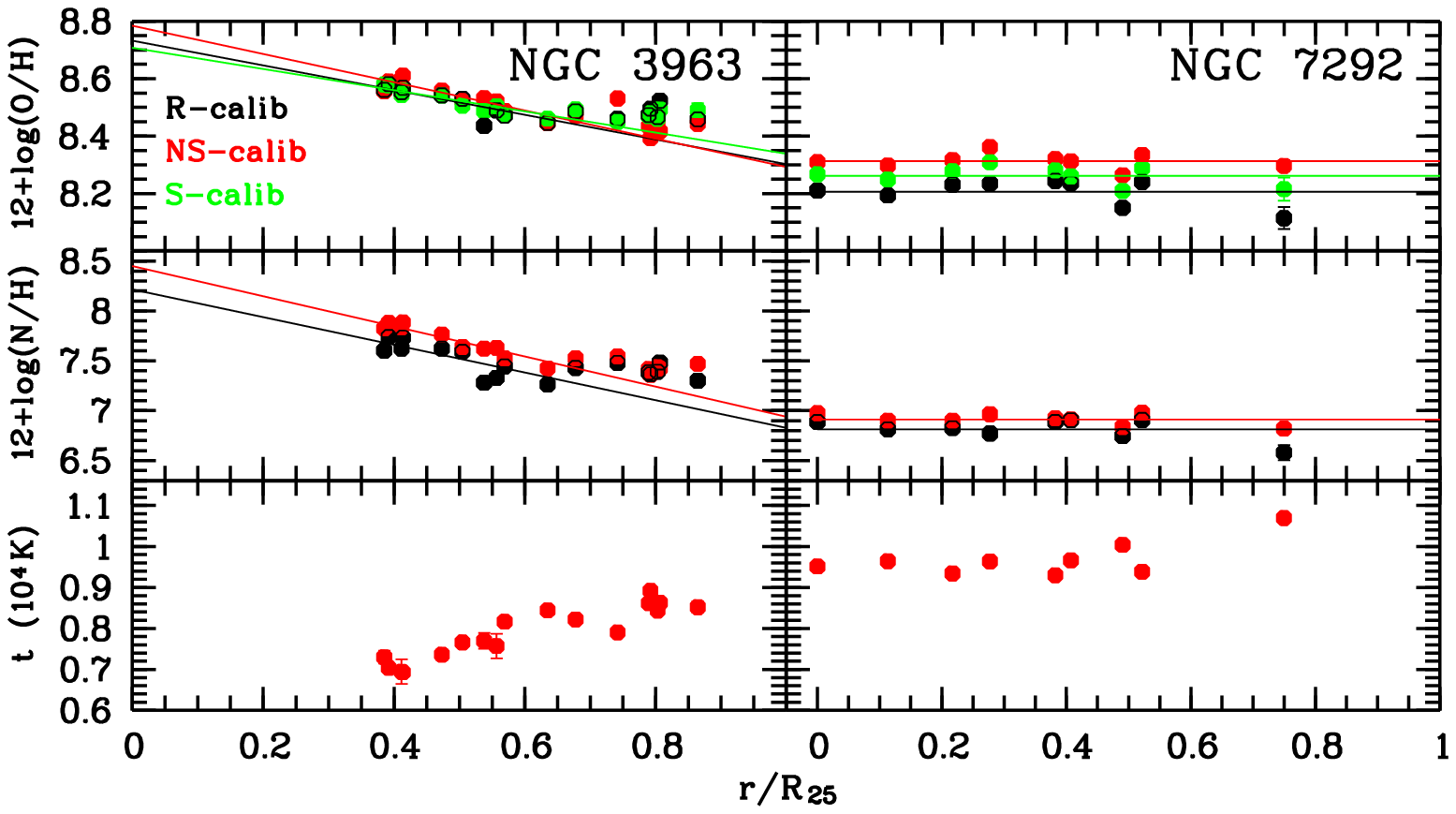}}
\caption{Radial distributions of oxygen abundances (top panels), 
nitrogen abundances (middle panels), and electron temperatures 
(bottom panels) in NGC~3963 (left) and NGC~7292 (right) calculated 
using R- (black), NS- (red), and S-calibrations (green). The solid 
lines are the best fits to data for NGC~3963 at $r<0.7R_{25}$ and 
NGC~7292. The error bars are shown. See the text for details.
}
\label{figure:grad}
\end{figure*}

\begin{table*}
\caption[]{\label{table:grad}
Parameters of radial distributions of oxygen and nitrogen 
abundances in the galaxies.
}
\begin{center}
\begin{tabular}{ccccccc} \hline \hline
Calibration & \multicolumn{3}{c}{$12+\log$(O/H)} & 
\multicolumn{3}{c}{$12+\log$(N/H)} \\
 & centre & \multicolumn{2}{c}{gradient} & centre & 
\multicolumn{2}{c}{gradient} \\
 & & (dex\,$R_{25}^{-1}$) & (dex\,kpc$^{-1}$) & 
   & (dex\,$R_{25}^{-1}$) & (dex\,kpc$^{-1}$) \\
\hline
\multicolumn{7}{c}{NGC~3963 ($r<0.7R_{25}$)} \\
R  & $8.73\pm0.09$ & $-0.43\pm0.09$ & $-0.023\pm0.005$ & 
$8.22\pm0.18$ & $-1.39\pm0.35$ & $-0.076\pm0.019$ \\
NS & $8.78\pm0.04$ & $-0.49\pm0.07$ & $-0.027\pm0.004$ & 
$8.45\pm0.09$ & $-1.51\pm0.18$ & $-0.082\pm0.010$\\
S  & $8.71\pm0.03$ & $-0.37\pm0.06$ & $-0.020\pm0.003$ &
$-$ & $-$ & $-$ \\
\multicolumn{7}{c}{NGC~3963 ($r\ge0.7R_{25}$)} \\
R  & $8.48\pm0.03$ & $0$ & $0$ & $7.40\pm0.07$ & $0$ & $0$ \\
NS & $8.44\pm0.05$ & $0$ & $0$ & $7.44\pm0.06$ & $0$ & $0$ \\
S  & $8.47\pm0.02$ & $0$ & $0$ & $-$ & $-$ & $-$ \\
\multicolumn{7}{c}{NGC~7292} \\
R  & $8.21\pm0.05$ & $0$ & $0$ & $6.81\pm0.11$ & $0$ & $0$ \\
NS & $8.31\pm0.03$ & $0$ & $0$ & $6.91\pm0.06$ & $0$ & $0$ \\
S  & $8.26\pm0.03$ & $0$ & $-$ & $-$ & $-$ & $-$ \\
\hline
\end{tabular}
\end{center}
\end{table*}

Radial distributions of oxygen and nitrogen abundances, as well as 
electron temperatures, obtained using three different 
calibrations, are presented in Fig.~\ref{figure:grad}.

Both galaxies show a peculiar radial abundance distribution. Both 
O/H and N/H decrease with distance from the galactic center to distances 
of $r\approx0.7R_{25}$ (13~kpc) in NGC~3963 (Fig.~\ref{figure:grad}), 
which is common for spiral galaxies \citep{pilyugin2014}. However, we 
find a flat distribution of oxygen and nitrogen abundances beyond these 
distances, at $r = 0.7-0.9R_{25}$. The flattening of radial oxygen 
abundance gradients in the outer parts of the discs in some giant 
galaxies was observed earlier. \citet*{ferguson1998} found it in 
NGC~628 at galactocentic distances $r>R_{25}$. Other similar examples are 
M83 and NGC~4625 \citep*{pilyugin2012}. However, this is a sufficiently 
rare occurrence. Among a sample of 130 nearby late type galaxies 
from \citet{pilyugin2014}, we found only three galaxies with 
the steep inner ($r<0.7-0.8R_{25}$) and flat outer distribution: 
NGC~1365, NGC~3621, and NGC~5457. All of them have signs of asymmetry in 
structure.

In the case of NGC~7292, we did not find any significant changes in 
oxygen and nitrogen abundances with distance from the center 
(Fig.~\ref{figure:grad}). Only one, the most distant H\,{\sc ii} region 
in this galaxy, is distinguished for its lower O/H and N/H abundances.

These peculiarities in chemical abundance distributions in both 
galaxies will be discussed in Section~\ref{sect:discus} in more detail.

To estimate radial oxygen and nitrogen abundances gradients, we used 
standard equations:
\begin{eqnarray}
12+\log({\rm O/H})=12+\log({\rm O/H})_0+C_{\rm O/H}r \nonumber
\end{eqnarray}
and 
\begin{eqnarray}
12 + \log({\rm N/H})=12+\log({\rm N/H})_0+C_{\rm N/H}r, \nonumber
\end{eqnarray}
where $12+\log({\rm O/H})_0$, $12+\log({\rm N/H})_0$ are the extrapolated 
central oxygen and nitrogen abundances,
$C_{\rm O/H}$, $C_{\rm N/H}$ are the slopes of the oxygen and nitrogen 
abundances gradients, and $r$ is the galactocentric distance.

The numerical values of the coefficients in the equations 
have been derived through the least squares method.

For oxygen and nitrogen abundances in NGC~7292 as well 
as in the outer disc of NGC~3963 ($r\ge0.7R_{25}$) we give 
the mean values.

Parameters of radial distributions of oxygen and nitrogen 
abundances in NGC~3963 and NGC~7292 are presented in 
Table~\ref{table:grad}.

\begin{figure*}
\vspace{1cm}
\resizebox{0.85\hsize}{!}{\includegraphics[angle=000]{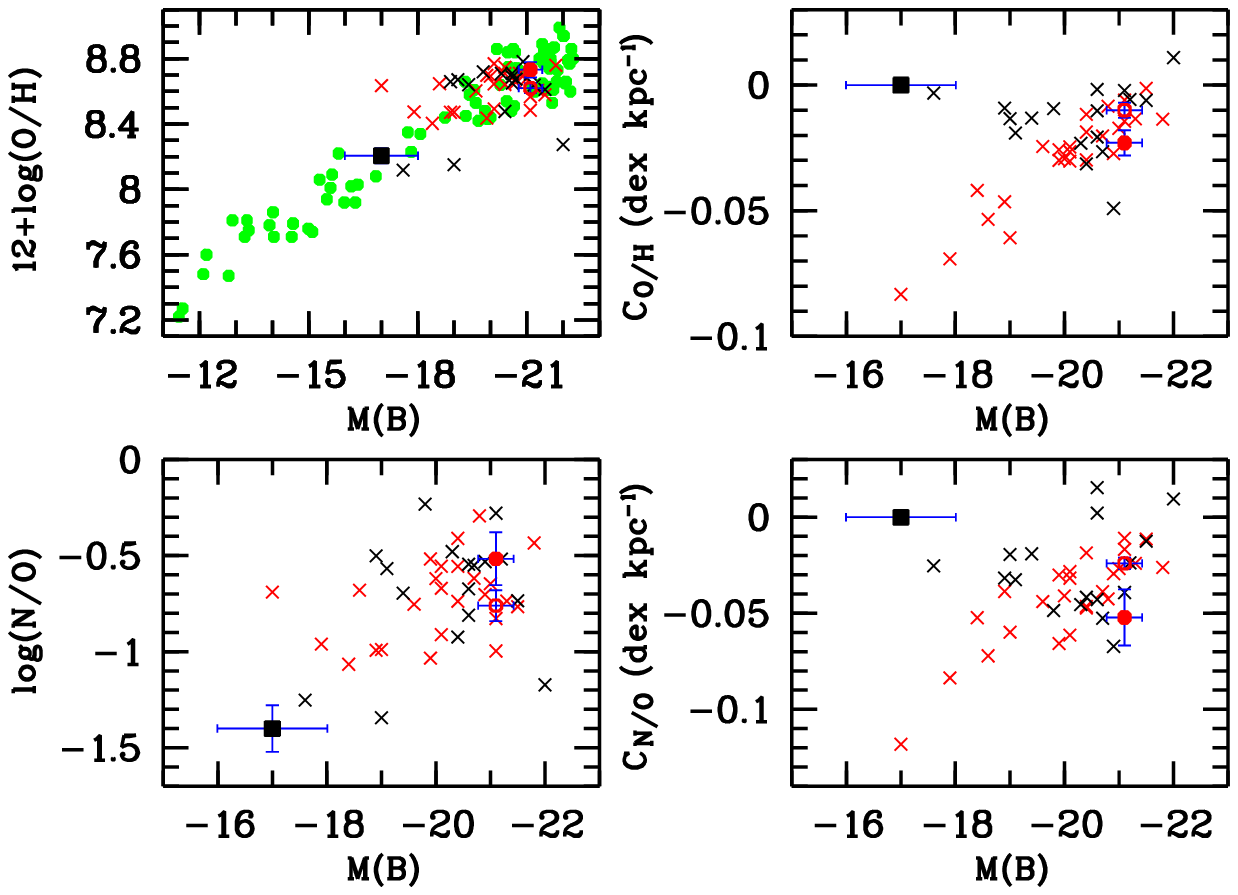}}
\caption{Central (mean) O/H and N/O and their gradients for galaxies 
of different luminosity. The green circles denote the central oxygen 
abundances in the discs of spiral galaxies and the oxygen abundances 
in irregular galaxies from \citet{pilyugin2007} with additional data 
from \citet{gusev2012}  and \citet{gusev2013}. The crosses indicate 
the parameters of strongly barred (black) and unbarred and weakly 
barred galaxies from \citet{zurita2021} obtained using the R-calibration. 
The filled red circle shows the parameters derived for H\,{\sc ii} 
regions sample in the inner disc of NGC~3963 ($r<0.7R_{25}$), the 
open red circle denotes the parameters of fitting of full sample 
of H\,{\sc ii} regions in NGC~3963, and the black square indicate 
the mean parameters of chemical abundance distribution in NGC~7292 
calculated using the R-calibration. The error crosses are shown. See 
the text for details.
}
\label{figure:mb}
\end{figure*}

Additionally, we calculated the central nitrogen-to-oxygen ratios, 
$\log({\rm N/O})_0$, and the slopes of N/O gradient, 
$C_{\rm N/O}$, using the equation
\begin{eqnarray}
\log({\rm N/O})=\log({\rm N/O})_0+C_{\rm N/O}r. \nonumber
\end{eqnarray}

Figure~\ref{figure:grad} and Table~\ref{table:grad} show that the 
radial O/H and N/H gradients, obtained using different calibrations, 
coincide within slope errors. Both oxygen and nitrogen 
gradients in the inner part of NGC~3963 are typical for giant 
spiral galaxies \citep[see e.g.][]{pilyugin2014,zurita2021b}.

Electron temperatures of H\,{\sc ii} regions anticorrelate with O/H 
and N/H abundances (Fig.~\ref{figure:grad}). This anticorrelation, 
indicating that the electron temperature in the nebula essentially 
depends on the cooling of gas through radiation in the nebular lines, 
is well known 
\citep[see the review in][for more details]{ellison2008,lopezsanchez2010}. 
However, it should be noted that O/H, N/H and $t_e$ are dependent 
parameters in empirical calibrations. Therefore, we can not 
investigate separately any temperature features in studied 
H\,{\sc ii} regions.

\section{Discussion}
\label{sect:discus}

\subsection{General chemical distribution parameters}

In Fig.~\ref{figure:mb} we compared general parameters of the chemical 
elements abundances distribution in NGC ~3963 and NGC~7292 (central (mean) 
O/H and N/O, and their gradients) with similar parameters of galaxies of 
different luminosity and morphology known from the literature.

\citet{zurita2021b} found that the O/H and N/O gradients -- luminosity 
relations demonstrate different behaviour for strongly barred and 
unbarred galaxies; strongly barred galaxies show shallow O/H and N/O 
gradients, whereas these gradients in unbarred and weakly barred galaxies 
become steeper as the galactic luminosity decreases.

Inspection of Fig.~\ref{figure:mb} reveals that both NGC~3963 and NGC~7292 
follow well the general trends in the diagrams. The giant weakly barred 
galaxy NGC~3963 lies within the spread of values of \citet{zurita2021} 
on all diagrams, although it has slightly steeper O/H and N/O gradients 
(see the filled red circles in Fig.~\ref{figure:mb}). Note, if we take 
formal results of linear fitting for the full sample of 
H\,{\sc ii} regions in NGC~3963 (open red circles in Fig.~\ref{figure:mb}), 
we will find that the galaxy is not distinguished by central O/H and 
N/O values and their linear gradients from those of sample of 
\citet{zurita2021}.

The strongly barred galaxy NGC~7292 belongs to a rare type of galaxies (see 
Introduction). Data on the abundance of chemical elements have so far been 
obtained only for a limited number of Magellanic-type galaxies (see e.g. 
the sample of \citet{pilyugin2007} in top-left panel of 
Fig.~\ref{figure:mb}). Among the galaxies studied by \citet{zurita2021}, 
only one galaxy, NGC~4395, is close to NGC~7292 in morphology and 
luminosity. As seen from Fig.~\ref{figure:mb}, NGC~7272 and NGC~4395 have 
similar central oxygen abundances, central oxygen-to-nitrogen ratios, 
as well as O/H and N/O gradients.

\subsection{NGC~3963 features}
\label{sect:discus2}

In spite of ordinary general chemical parameters, H\,{\sc ii} regions in 
the outer disc of NGC~3963 show the same oxygen and nitrogen abundances 
within the accuracy of the R-calibration independently of their galactocentic 
distance (Fig.~\ref{figure:grad}, see also the middle part of 
Table~\ref{table:grad}). Moreover, their O/H and N/H abundances, and N/O 
ratio seem to be slightly higher than those in H\,{\sc ii} regions from the 
middle part ($r\approx0.55R_{25}$) of the NGC~3963 disc. All these H\,{\sc ii} 
regions (nos.~17-20, 22, 23) are located in the southern and south-western 
parts of NGC~3963 (Fig.~\ref{figure:map}). The deprojected linear distance 
between the outermost regions (nos.~17 and 23) is 13~kpc. Recall that the 
objects nos.~17-20 were observed using TDS (slit position~3) while the 
regions nos.~22, 23 were observed using BOSS, i.e. in two 
independent projects.

\begin{figure}
\vspace{0.8cm}
\hspace{0.7cm}
\resizebox{0.80\hsize}{!}{\includegraphics[angle=000]{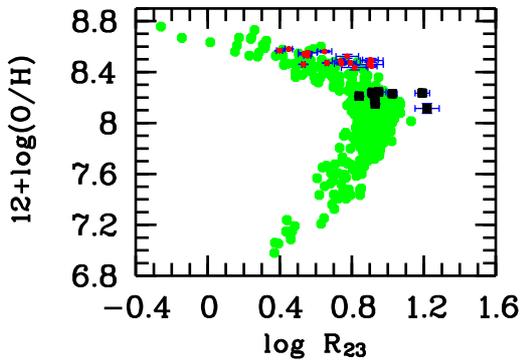}}
\caption{The $\log R_{23}-$O/H diagram for the sample of H\,{\sc ii} 
regions investigated in the present study (symbols are the same as 
in Fig.~\ref{figure:nhoh}) and taken from the literature 
\citep{pilyugin2012,gusev2012,gusev2013}.
}
\label{figure:r23}
\end{figure}

As we noted in the Introduction, NGC~3963 has a companion, NGC~3958, which is 
located at the distance of 110~kpc from NGC~3963 in the SSW direction. This 
pair has been repeatedly studied in the 21~cm line 
\citep{moorsel1983,nordgren1997}. Results of H\,{\sc i} data analysis 
strongly suggest the presence of tidal distortion in both galaxies in 
the pair \citep{moorsel1983}. Actually, features of the H\,{\sc i} spatial 
distribution and the H\,{\sc i} velocity field in NGC~3963 can be clearly 
interpreted as the presence of a powerful gas inflow from the south-western 
direction \citep[see H\,{\sc i} maps and velocity fields for NGC~3963 
in][]{moorsel1983,nordgren1997}. This flow shoves gas behind the nucleus 
into the north-eastern part of the galaxy.

We assume that the inflow of metal-enriched gas into the south-western part 
of the galaxy can stimulate star formation in the southern and south-western 
parts of the outer disc of NGC~3963. As a result, we observe H\,{\sc ii} 
regions with the same, comparatively rich chemical abundance. It is 
possible that this stream is also responsible for the distortion of the 
southern spiral arm from the form of the classical logarithmic spiral 
(see Fig.~\ref{figure:map}).

We do not discuss H\,{\sc ii} region no.~14 with $r=0.68R_{25}$, 
which is located at the end of the northern spiral arm of NGC~3963 
(Fig.~\ref{figure:map}). It has intermediate O/H and N/H abundances 
between the regions in the inner and the outer discs (Fig.~\ref{figure:grad}), 
but we have not any other spectral data for H\,{\sc ii} regions in the 
outer disc in the northern part of NGC~3963.

\begin{figure}
\vspace{0.6cm}
\resizebox{1.00\hsize}{!}{\includegraphics[angle=000]{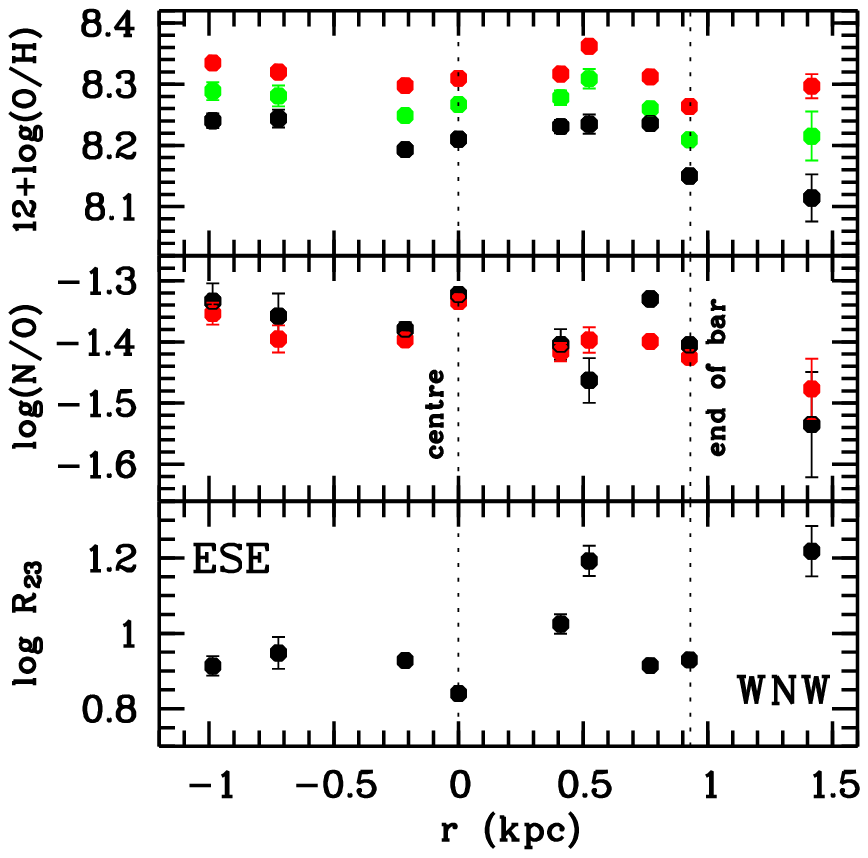}}
\caption{Distributions of oxygen abundance (top), nitrogen-to-oxygen 
ratio (middle), and $\log R_{23}$ (bottom) along the bar of NGC~7292. 
Symbols are the same as in Fig.~\ref{figure:grad}. The centre and the 
end of the bar of the galaxy are indicated by dotted lines. See the text 
for details.
}
\label{figure:bar}
\end{figure}

\subsection{NGC~7292 features}
\label{sect:discus3}

Oxygen and nitrogen abundances usually do not decrease with 
galactocentric distance in dwarf and irregular galaxies 
\citep*{richer1995,miller1996,pilyugin2001,hidalgo2001,
testor2001,testor2003,peimbert2005,hernandez2009}. NGC~7292 is a good 
example of an irregular galaxy without radial abundance gradient. However, 
the chemical elements abundance distribution in NGC~7292 is neither 
strongly constant nor chaotic.

We noted in Section~\ref{sect:pre_res} H\,{\sc ii} regions nos.~1 and 4 
which are located near the border separating objects with thermal 
and non-thermal emissions on the [N\,{\sc ii}]$\lambda$6584/H$\alpha$ 
versus [O\,{\sc iii}]$\lambda$5007/H$\beta$ diagnostic diagram 
(Fig.~\ref{figure:nhoh}). Both these regions stand out in the 
$\log R_{23}-$O/H diagram (Fig.~\ref{figure:r23}) too.

This diagram, where
\begin{center}
$R_{23} = I$(([O\,{\sc ii}]$\lambda$3727+3729$+$
[O\,{\sc iii}]$\lambda$4959$+$[O\,{\sc iii}]$\lambda$5007)/$I$(H$\beta$)), 
\end{center}
separates cool rich-oxygen and hot low-oxygen H\,{\sc ii} regions 
\citep[see e.g.][]{pilyugin2010}. If the objects from NGC~3963 occupate 
the area, where cool rich-oxygen regions, typical for giant galaxies, 
are located, while most objects from NGC~7292 occupate the area, where 
warm moderate-metallicity H\,{\sc ii} regions are located, then 
regions nos.~1 and 4 from NGC~7292 show extremely high $R_{23}$ for any 
O/H (Fig.~\ref{figure:r23}). This indicates a significant role of 
the non-thermal (shock) emission in the formation of spectra of these 
H\,{\sc ii} regions. Regions nos.~1 and 4 are also distinguished by the 
maximum errors in O/H, N/H, and N/O estimates and the largest differences 
between O/H, N/H, and N/O values calculated using different calibrations 
(see Figs.~\ref{figure:grad}, \ref{figure:bar}, Table~\ref{table:abun}). 
It is interesting that the region no.~4 is located in the centre of the bar 
of NGC~7292 at 500~pc from the nucleus and the faint region no.~1 is located 
in the outer western part of the galaxy (Fig.~\ref{figure:map}).

We have considered the O/H, N/O, and $\log R_{23}$ distributions along 
the bar (and the major axis) of NGC~7292 in Fig.~\ref{figure:bar}.

The nucleus (the region no.~6) of NGC~7292 shows the highest N/O ratio among  
H\,{\sc ii} regions in the galaxy. The second by brightness region no.~2 
(the end of the bar; see Fig.~\ref{figure:map}) has minimum oxygen 
abundance and low N/O ratio (Fig.~\ref{figure:bar}). In general, we 
observe a weak N/O gradient along the major axis of NGC~7292: the N/O ratio 
decreases from the eastern to the western part of the galaxy (see the 
middle panel in Fig.~\ref{figure:bar}).

Higher values of N/O correspond to smaller specific star formation 
rate within an H\,{\sc ii} region, i.e. star formation in it could have 
been high in the past, at earlier stage of evolution 
\citep{mallery2007,molla2010}. The lower N/O ratios correspond to 
younger, less evolved H\,{\sc ii} regions. Thus, the observed N/O 
gradient may indicate the propagation of a star formation wave along 
the major axis of NGC~7292 from the east to the west of the galaxy. 
Unfortunately, the absence of H\,{\sc i} and H\,{\sc ii} velocity 
fields makes it impossible to develop our hypothesis without additional 
observational data.

\section{Conclusions}

The  spectroscopic observations of 32 H\,{\sc ii} regions in NGC~3963 
and NGC~7292, obtained with the 2.5-m telescope of the Caucasus Mountain 
Observatory of the Sternberg Astronomical Institute with the newly 
Transient Double-beam Spectrograph, were carried out.

The oxygen and nitrogen abundances in 26 H\,{\sc ii} regions are 
estimated using different empirical calibrations. Parameters 
of radial distributions of the oxygen and nitrogen abundances in 
the galaxies are obtained. The chemical elements abundance data 
for H\,{\sc ii} regions in NGC~3963 and NGC~7292 are estimated for 
the first time.

General chemical elements abundances and their distribution in 
NGC~3963 and NGC~7292 are typical for galaxies with similar 
luminosities and morphology. Wherein, both galaxies have some 
peculiarities in chemical abundance distributions.

H\,{\sc ii} regions in the outer part of southern spiral arm of 
NGC~3963 at distances $r>0.7R_{25}$ show constant, slightly higher 
oxygen and nitrogen abundances. This can be explained by the inflow 
of metal-enriched gas into the south-western part of NGC~3963.

Oxygen and nitrogen abundances do not decrease with galactocentric 
distance in NGC~7292. However, the nitrogen-to-oxygen ratio seems to 
decrease along the major axis from the eastern to the western part of the 
galaxy.

\section*{Acknowledgments}
We are grateful to M.~A.~Burlak (SAI MSU) for help and support during 
the observations, to O.~V.~Egorov (Astronomisches Rechen-Institut, 
Universit\"{a}t Heidelberg and SAI MSU) for helpful comments on 
different empirical calibrations, and to E.~V.~Shimanovskaya (SAI MSU) 
for help with editing this paper. This study was supported by the 
Russian Foundation for Basic Research (project no~20-02-00080). 
AVD acknowledges the support by the Interdisciplinary Scientific and 
Educational School of Moscow University ''Fundamental and Applied 
Space Research''. The authors acknowledge support from M.V.~Lomonosov 
Moscow State University Program of Development in expanding the 
instrumentation base of the CMO SAI MSU. The authors acknowledge the 
use of the HyperLeda data base (\url{http://leda.univ-lyon1.fr}), 
the NASA/IPAC Extragalactic Database 
(\url{http://ned.ipac.caltech.edu}), and The Sloan Digital Sky 
Survey (\url{http://www.sdss.org}).

\section*{Data availability}
The SDSS data used in this article are available in the 
SDSS-DR16 database at http://skyserver.sdss.org/dr16/. The TDS 
spectroscopic data can be shared on reasonable request to the 
corresponding author.

\end{document}